\documentclass[journal]{IEEEtran}

\IEEEoverridecommandlockouts

\usepackage{caption}
\usepackage{subcaption}
\usepackage{graphicx}
\usepackage{cite}
\usepackage{url}
\usepackage[cmex10]{amsmath}
\usepackage{amssymb}
\usepackage{algorithm}
\usepackage{algpseudocode}
\usepackage{cases}
\usepackage{color}
\usepackage{soul}
\usepackage{xcolor}
\usepackage{framed}
\usepackage{cite}
\usepackage{epstopdf}
\usepackage{calc}
\usepackage{array}
\usepackage{amsmath}
\usepackage{comment}
\usepackage{dsfont}
\DeclareGraphicsExtensions{.pdf,.eps,.png,.jpg,.mps}
\begin{document}
\font\myfont=cmr12 at 20pt
\title{\huge Kuiper Test based Modulation Level Classification under Unknown Frequency Selective Channels}
\author{\IEEEauthorblockN{Shailesh Chaudhari and Danijela Cabric\\}
\IEEEauthorblockN{Department of Electrical Engineering,
University of California, Los Angeles\\
Email: schaudhari@ucla.edu, danijela@ee.ucla.edu}
\vspace{-7mm}}

\maketitle


\begin{abstract}
	In this paper, we address the problem of identifying the modulation level of the received signal under an unknown frequency selective channel. The modulation level classification is performed using reduced-complexity Kuiper (rcK) test which utilizes the distribution of signal features such as magnitude of the received samples or phase difference in consecutive received samples. However, in frequency selective channels, these features are severely distorted resulting in a poor classification performance. We propose to use constant modulus algorithm (CMA) to mitigate the impact of the frequency selective channel on the signal feature. Simulation and analytical results show that the proposed CMA-rcK technique outperforms state-of-the-art cumulant-based technique as well as blind equalizer-based technique that uses Alphabet Matched Algorithm. 
\end{abstract}

\IEEEpeerreviewmaketitle

\begin{IEEEkeywords}
Alphabet matched algorithm, constant modulus algorithm, reduced-complexity Kuiper, cumulants.
\vspace{-2mm}
\end{IEEEkeywords}


\section{Introduction}
\label{sec:Introduction}

Modulation level classification is a process of identifying the modulation level (order) of the received signal of a known modulation type \cite{urriza2011}. For example, if the received signal is known to be of QAM modulation type, then the modulation level classification identifies the modulation order of the QAM signal, e.g., 4-/16-/64-QAM. Modulation level classification can be used in order to identify interference signals or to identify licensed primary transmitter in a cognitive radio network \cite{ramkumar2009}. The level classification information can also be used in military applications for surveillance of unknown signals as well as in communication systems employing adaptive modulation schemes\cite{ramkumar2009, ding2016}.

Modulation level classification techniques can be broadly divided into two classes: likelihood-based and feature-based \cite{dobre2007}. Likelihood-based methods, although optimal in Bayesian sense, suffer from high computational complexity. The feature-based methods include classification using cumulants \cite{swami2000, wu2008, orlic2010} as well as  goodness-of-fit test of distribution of a feature of the signal \cite{wang2010, wang2016, urriza2011}. Magnitude of the signal is used as a feature to classify levels of QAM and PAM signals, while the phase difference between consecutive samples is used as a feature to classify levels of PSK signals. The distance between empirical cumulative distribution function (ECDF) and theoretical CDF is used to identify the modulation level. The work in \cite{wang2010} proposed Kolmogorov-Smirnov (KS) test to compute the distribution distance and classify the modulation level. A fold-based Kolmogorov-Smirnov classifier is proposed in \cite{wang2016} to improve the performance of the KS classifier. A computationally efficient method, called reduced complexity Kuiper (rcK) was proposed in \cite{urriza2011} that computes the distribution distance using a finite set of testpoints where the distribution distance between features of two modulation levels is maximum.

The classification techniques in \cite{swami2000, wang2010, wang2016, urriza2011} do not consider the impact of the frequency selective channel. The works in \cite{wu2008, orlic2010, liu2014a, marey2014, barbarossa2000} have developed classification techniques under frequency selective channels. However, these techniques have the following shortcomings. The classifiers in \cite{wu2008, orlic2010, liu2014a} utilize fourth- and sixth-order cumulants in order to classify between 4-, 16-, and 64-QAM. These cumulants cannot be used to classify between higher order M-PSK ($M>2$) signals  because the cumulant values for those modulations are the same \cite{dobre2007}. Further, the work in \cite{marey2014} proposed to use higher order moments to identify modulation of the received signal under frequency selective channel. This method can be used for binary classification between a modulation types for which the moment is zero and non-zero. Therefore, this method cannot be used to identify modulation levels of higher order M-QAM ($M>4$) and M-PSK ($M>2$). The works in \cite{swami2000a, barbarossa2000} have proposed an adaptive blind equalizer-based approach using alphabet matched algorithm (AMA). In this technique, the modulation level is decided based on the closest match between equalized symbols and constellation points of modulation levels. This technique, however, performs poorly at SNRs below 10dB \cite{hatzichristos2001}. 

In this paper, we propose a new technique for modulation level classification under frequency selective channels that employs constant modulus algorithm (CMA) to blindly equalize the received symbols. The output of the equalizer is provided to reduced complexity Kuiper (rcK) classifier, which outputs the modulation level based on goodness-of-fit test. We also provide expression for the variance of error at the output of the equalizer. This expression is used to analytically compute the probability of correct classification.

The paper is organized as follows. The system model is presented in Section \ref{sec:Model} along with the proposed CMA-rcK method. The expression for the variance of error at the output of CMA equalizer is provided in Section \ref{sec:cma_analysis}. Simulation results are presented in Section \ref{sec:Results} to show the performance of classification techniques under various frequency selective channels. Finally, the concluding remarks are provided in Section \ref{sec:Conclusion}.

\textit{Notations}: In this paper, lowercase and uppercase bold letters indicate column vectors and matrices, respectively, e.g., $\mathbf{x, H}$. Scalar variables are denoted by non-bold letters e.g., $s,v$. The modulation level (order) is denoted by $\mathcal{M}_k$. Notations $(.)^T$, and $(.)^H$ denote transpose and Hermitian operations.
\section{System Model and CMA-rcK algorithm}
\label{sec:Model}
Let $c_k(i), i =1,2,...,\mathcal{M}_k$ be the set of constellation points of a modulation order $\mathcal{M}_k \in \{\mathcal{M}_1, \mathcal{M}_2,..., \mathcal{M}_K\}$, where $K$ is the total number modulation levels to be classified. Let $s_k$ be the transmitted baseband symbols randomly selected from $c_k$. The symbols are transmitted over a multipath channel with $Q$ taps: $\mathbf{h}=[h(0),h(2),...,h(Q-1)]^T$. We consider the following model for the received signal \cite{wu2008, marey2014, barbarossa2000}:
\begin{align}
x(n)= \sum_{q=0}^{Q-1}h(q)s_k(n-q)+v(n),
\label{eq:rx_signal}
\end{align}
where $v(n)\sim \mathcal{CN}(0,\sigma^2_v)$ is additive white Gaussian noise (AWGN). The noise is assumed to be independent of transmitted symbols $s_k$ and the noise variance $\sigma^2_v$ is assumed to be known. The transmitted symbols $s_k's$ are zero mean and are normalized to have unit power, $\mathbb{E}[|s_k|^2]=1$ . We denote the received noise-free signal by $x'(n)=\sum_{q=0}^{Q-1}h(q)s_k(n-q)$.

\subsection{Proposed CMA-rcK technique for level classification}
\label{sec:Cma_rcK}
In order to classify the modulation level, the received signal $x(n)$ is passed through a CMA equalizer of length $L$. Let $\mathbf{x}_i = [x(iL-1),x(iL-2),...,x(iL-L)]^T$ be $L\times 1$ input vector to the equalizer for the $i$-th tap update. It should be noted that we are considering block processing in this paper. Therefore, all of the input samples in the vectors $\mathbf{x_{i}}$ and $\mathbf{x_{i+1}}$ are different. The output of the equalizer after the $i$-th tap update is $y(i)=\mathbf{w}^H_{i-1} \mathbf{x}_i $, where $\mathbf{w}_i=[w_i(0),...,w_i(L-1)]^T$ is the equalizer tap vector after the $i$-th update. The CMA equalizer taps are adapted in oder to minimize the cost function $J_{cma} = \mathbb{E}[(|y(i)|^2 -R)^2]$, where $R$ is a CMA parameter. The value of parameter $R=\frac{\mathbb{E}[|s_k|^4]}{\mathbb{E}[|s_k|^2]}$ in the above equation depends on the transmitted modulation level \cite{johnson1998}. Since the modulation level is unknown, we set $R=1$, which is the parameter value for constant modulus signals such as 4-QAM and M-PSK. It should be noted that even though the equalizer name includes constant modulus, it can be used to equalize the channel even when the transmitted symbols are of higher order QAM signals \cite{johnson1998}. The tap update equation for the CMA is given as follows:
\begin{align}
\mathbf{w}_i=\mathbf{w}_{i-1} - \mu (|y(i)|^2-1)y^{*}(i)\mathbf{x}_i,
\label{eq:cma_update}
\end{align}
where $\mu$ is the step size. The equalizer output after $M$ iterations is denoted by $y_{eq}(i)=\mathbf{w}^H_M \mathbf{x}_i$. The output $y_{eq}(i)$ is passed to rcK classifier \cite{urriza2011} for classification of the modulation level, $\mathcal{M}_{\hat k} \in \{\mathcal{M}_{1},...,\mathcal{M}_{K}\}$. 

In order to classify the modulation level of $y_{eq}(i)$, the rcK block uses the estimated SNR at the output of equalizer $\hat \gamma= \frac{\frac{1}{M}\sum_i |y_{eq}(i)|^2 }{||\mathbf{w}_M||^2 \sigma^2_v} -1$. The rcK classifier uses feature of the equalized symbols, denoted by $f(y_{eq})$, in order to identify the modulation level. For QAM signals,  the magnitude feature is used, i.e., $f(y_{eq})=|y_{eq}(i)|$, while for PSK signals the phase difference between consecutive symbols is used as a feature, i.e., $f(y_{eq})=\angle y_{eq}(i) -\angle y_{eq}(i-1)$ \cite{urriza2011,wang2010}. 

The rcK classifier computes the Kuiper distance $V_l$ between the empirical cumulative distribution function (ECDF) of feature of $y_{eq}$ and theoretical CDF of $f(s_l+g)$, where $l=1,2,...,K$ and $g \sim \mathcal{CN}(0,\sigma^2_g)$ is AWGN noise \cite{urriza2011}. The noise variance $\sigma^2_g$ is set such that $\hat \gamma = 1/\sigma^2_g$. Let $F_{y_{eq}}(t)$ be the ECDF of the equalized symbols defined as
\begin{align}
F_{y_{eq}}(t) = \frac{1}{M}\sum_{m=1}^{M}\mathbb{I}(f(y_{eq})\leq t),
\label{eq:ecdf}
\end{align}
where $\mathbb{I}(.)$ equals to one if the input is true, and zero otherwise. Similarly, we define $F_0^l(\tau)$ as the theoretical CDF of the feature $f(s_l+g)$: $F_0^l(\tau)= Pr(f(s_l+g) \leq \tau)$. The Kuiper distance $V_l$ is the distance between the ECDF and the CDF computed at predefined testpoints $t_{lp}^{(\delta)}$ where the positive and negative deviation between the CDFs $F_0^l(\tau)$ and $F_0^p(\tau)$ is maximum \cite{urriza2011}. The testpoints for levels $l$ and $p$ are obtained as:
\begin{align}
t_{lp}^{(\delta)} = \arg\max_{\tau} (-1)^{\delta} \left(F^l_0(\tau) - F^p_0(\tau)\right),
\end{align}
for $l,p\in [1,K]$, and $\delta \in \{0,1\}$. Note that $\delta=0$ corresponds to positive deviation while $\delta=1$ corresponds to negative deviation. The testpoints $t_{lp}^{\delta}$ are obtained offline using CDFs $F^l_0(\tau)$ \cite{wang2010}. We need to obtain two testpoints for each pair of $l$ and $p$, $l\neq p, l,p \in [1,L]$ for each value of $\gamma$. Therefore, the number of testpoints required per SNR value is equal to $2 \binom{L}{2}$.

The Kuiper distance between the ECDF $F_{y_{eq}}(t_{lp}^{(\delta)})$ and the CDF $F^l_0(t_{lp}^{(\delta)})$ is the sum of maximum positive and negative deviations, as follows
\begin{align}
{V}_l = |  D_l^{(0)} + D_l^{(1)}  |,
\end{align}
where  $D_l^{(\delta)} = (-1)^{(\delta)} \left(F_{y_{eq}} \left(t_{lp}^{(\delta)}\right)  -F^l_0 \left(t_{lp}^{(\delta)}\right)\right)$. The modulation level is classified as the level $l$, whose CDF closest to the ECDF in Kuiper distance sense:
\begin{align}
\hat k = \arg \min \limits_l V_l
\end{align}

\begin{algorithm}[ht]
	\caption{CMA-rcK algorithm}
	\label{alg:cma_rck}
	\begin{algorithmic}[1]
		\State Input: $x(n), n=1,2,\cdots,ML$, CMA initialization $\mathbf{w}_0=[1,0,0,..0]$, step size $\mu = 10^{-4}$, rcK testpoints $t_{lp}^{(\delta)}$.
		\While{$i\leq M$}
		\State {\small $\mathbf{x_i} = [x(iL-1),x(iL-2),...,x(iL-L)]^T$, $y(i)= \mathbf{w}_{i-1}\mathbf{x}_i$}
		\State CMA update: $\mathbf{w}_i=\mathbf{w}_{i-1} - \mu (|y(i)|^2-1)y^{*}(i)\mathbf{x}_i$
		\State $i=i+1$
		\EndWhile
		\State Equalization: $y_{eq}(i) = \mathbf{w}^H_M\mathbf{x}_i, i=1,2,\cdots, M$.
		\State SNR estimation: $\hat \gamma= \frac{\frac{1}{M}\sum_i |y_{eq}(i)|^2 }{||\mathbf{w}_M||^2 \sigma^2_v} -1$.
		\State Compute ECDF $F_{y_{eq}}(t_{lp}^{(\delta)})$.
		\State RcK level classification: $\hat{k} = \arg\min_{l} V_l, l=1,\cdots,K$.
	\end{algorithmic}
\end{algorithm}

The proposed CMA-rcK algorithm is summarized in Algorithm \ref{alg:cma_rck}. The total number of received samples used by the proposed CMA-rcK classifier is $ML$. The CMA block uses $ML$ samples of $x(n)$ for equalization, while the rcK block uses $M$ samples of $y_{eq}(i)$.

\section{Performance Analysis of CMA-rcK }
\label{sec:cma_analysis}
The probability of classifying the modulation level as $\mathcal{M}_l$ under channel $\mathbf{h}$, when the modulation level $\mathcal{M}_k$ is transmitted, is denoted by $\Pr(\mathcal{M}_{\hat k} = \mathcal{M}_l | \mathcal{M}_k, \mathbf{h})$. 
In order to evaluate $\Pr(\mathcal{M}_{\hat k} = \mathcal{M}_l | \mathcal{M}_k, \mathbf{h})$, we first model the distribution $F_{y_{eq}}$. The equalized symbols $y_{eq}$ at the output of CMA include the effect of residual equalizer error, called excess mean square error (EMSE), as well as noise enhancement due to equalization. We model the distribution $F_{y_{eq}}$ as the distribution of $f(s_k+\epsilon)$, where $\epsilon \sim \mathcal{CN}(0,\sigma^2_\epsilon)$ is modeled as a Gaussian random variable incorporating the effects of residual error and noise enhancement \cite{sayed2008}. Thus, the probability of correct classification using subsequent rcK classifier can be obtained in terms of SNR $\gamma = 1/\sigma^2_\epsilon$. The variable $\epsilon$ that indicates the error at output of the equalizer is denoted by
\begin{align}
\nonumber	\epsilon(i) = & y_{eq}(i) - s_k(iL-D)e^{j\theta}
= \mathbf{w}^H_M \mathbf{x}_i -s_k(iL-D)e^{j\theta}\\
= &\mathbf{w}^H_M \mathbf{x'}_i + \mathbf{w}^H_M \mathbf{v}_i-s_k(iL-D)e^{j\theta},
\label{eq:error}
\end{align}
where $D$ and $\theta$ are constant delay and phase, respectively. \footnote{\scriptsize $D$ and $\theta$ are the position of unit entry in $\mathbf{w^H_{zf} H}$ and corresponding phase, respectively, where $\mathbf{w}_{zf}$ are ZF equalizer taps and $\mathbf{H}$ is the Toeplitz matrix.} 
The vector  $\mathbf{v}_i=[v(iL-1),...,v(iL-L)]^T$ includes noise samples, while the vector $\mathbf{x'}_i = \mathbf{x}_i -\mathbf{v}_i $ includes noise-free received symbols. It should be noted that $\mathbb{E}[\epsilon]=0$, since $\mathbf{x_i}$ and $s_k$ are zero-mean variables. The consecutive $\epsilon(i)$ are independent due to block processing, since $\mathbf{x'}_i$ and $\mathbf{v}_i$ are independent of $\mathbf{x'}_{i-1}$ and $\mathbf{v}_{i-1}$. The variance $\sigma^2_\epsilon$ is
\begin{align}
\sigma^2_\epsilon =  \mathbb{E}[|\mathbf{w}^H_M \mathbf{x'}_i - s(iL-D)e^{j\theta}|^2] +  \mathbb{E}[|\mathbf{w}^H_M \mathbf{v}_i|^2].
\label{eq:error_var}
\end{align}
The first term in the above equations is the EMSE, the second term indicates noise enhancement due to equalizer taps. The analytical expression for EMSE of CMA is provided in \cite{sayed2008} under the condition that there exists a zero forcing (ZF) equalizer of length $L$ that completely eliminates ISI. However, the ZF equalizer, in general, has infinite impulse response (IIR). Let us denote the ZF equalizer for channel $\mathbf{h}$ by $\mathbf{w}_{zf} = \mathcal{Z}^{-1}\{1/H(z)\}$. Here, $H(z)=\sum_{q=0}^{Q-1}h(q)z^{-q}$ is the z-transform of the channel, while $\mathcal{Z}^{-1}$ indicates the inverse z-transform. Due to IIR nature of ZF equalizer, $\mathbf{w}_{zf} = [w_{zf}(0), w_{zf}(1), w_{zf}(2),...]^T$ has infinite taps in general. However, for practical purposes, we assume that the ZF equalizer can be approximated by an FIR filter with approximation length $L_{zf}$ taps. 

We assume that $w_{zf}(l)=0$ for $l\geq L_{zf}$, which makes $\mathbf{w}_{zf}$ a FIR filter with $L_{zf}$ taps: $\mathbf{w}_{zf}=[w_{zf}(0), w_{zf}(1),...,w_{zf}(L_{zf}-1)]^T$.  The approximation length $L_{zf}$ is larger if the channel is highly frequency selective with zeros of $H(z)$ closer to the unit circle as compared to a mildly frequency selective with the zeros closer to the origin. Therefore, under a highly frequency selective channel, we may have a case where CMA equalizer length $L$ is smaller than $L_{zf}$ and there is no ZF equalizer of length $L$ that can eliminate the ISI. Therefore, we need to take into account the residual ISI in order to compute the EMSE term.

Let us consider a truncated ZF equalizer of length $L$, denoted by $\mathbf{w}_{zf,L} = [w_{zf}(0), w_{zf}(1),...,w_{zf}(L-1)]^T$. Further, let $\mathbf{x'}_{i,zf}=[x'(iL-1),x'(iL-2),...x'(iL-L_{zf})]^T$ denote a $L_{zf}\times 1$ vector with all received samples in the absence of noise. The ZF response that completely eliminates ISI is $\mathbf{w}^H_{zf}\mathbf{x'}_{i,zf}=s(iL-D)e^{j\theta}$. This term can be rewritten in terms of the response of truncated ZF equalizer of length $L$ as $\mathbf{w}^H_{zf}\mathbf{x'}_{i,zf} = \mathbf{w}^H_{zf,L}\mathbf{x'}_i + \Delta(i)$, where $\Delta(i)=\sum_{l=L}^{L_{zf}-1}w_{zf}(l)x'(iL-l-1)$ is  residual ISI. 
The EMSE term in (\ref{eq:error_var}) can then be  written as follows:
{\small
	\begin{align}
	\nonumber &\mathbb{E}[|\mathbf{w}^H_M \mathbf{x'}_i - s(iL-D)e^{j\theta}|^2]
	= \mathbb{E}[|\mathbf{w}^H_M \mathbf{x'}_i - \mathbf{w}^H_{zf,L}\mathbf{x'}_i- \Delta(i)|^2]
	\\ &~~~~= \mathbb{E}[|\mathbf{w}^H_M \mathbf{x'}_i - \mathbf{w}^H_{zf,L}\mathbf{x'}_i^2] +   \mathbb{E}[|\Delta(i)|^2]
	\label{eq:emse}
	\end{align}}
The second equality follows from the fact that $\Delta(i)$ is independent of $\mathbf{x'}_i$ and $\mathbb{E}[\mathbf{x'}_i]=0$. The EMSE without residual ISI is given by \cite{sayed2008}:
	\begin{align}
	 \nonumber
	\mathbb{E}[|\mathbf{w}^H_M \mathbf{x'}_i &- \mathbf{w}^H_{zf,L}\mathbf{x'}_i^2] = 
	\\&\mu \frac{\mathbb{E}[|s_k|^6]-2 \mathbb{E}[|s_k|^4] + \mathbb{E}[|s_k|^2] }{4\mathbb{E}[|s_k|^2] -2 } \text{Tr}(\mathbf{HH^H}),
	\end{align}
	where $\mathbf{H}$ is a Toeplitz channel matrix of size $L \times (L+Q-1)$ with the first row $	[h(0), h(1), \cdots, h(Q-1),0,.. ]$.
Further, the second term in (\ref{eq:emse}) is the contribution of residual ISI due to truncated ZF equalizer and is given as $\mathbb{E}[|\Delta(i)|^2] = ||\mathbf{w'}^H_{zf,L} \mathbf{H'}||^2$, where $\mathbf{w'_{zf,L}}=[w_{zf}(L),w_{zf}(L+1),...,w_{zf}(L_{zf})]^T$ contains ZF taps not included the truncated ZF equalizer $\mathbf{w}_{zf,L}$, while $\mathbf{H'}$ is Toeplitz channel matrix of size $(L_{zf}-L+1) \times (L_{zf}-L+Q)$. Finally, the noise enhancement at the output of equalizer is given as \cite{fijalkow1997}: 
$\mathbb{E}[|\mathbf{w}^H_M \mathbf{v}_i|^2] = \sigma^2_v \mathbf{e}^H_D \mathbf{(HH^H)^{-1}} \mathbf{e}_D$, where $\mathbf{e}_D=[0,...,1,...,0]^T\in \mathbb{R}^{L\times 1}$, where $1$ is at $D$-th location. We obtain variance of error, $\sigma^2_\epsilon$ as follows:
	\begin{align}
	\nonumber \sigma^2_\epsilon =& \mu \frac{\mathbb{E}[|s_k|^6]-2  \mathbb{E}[|s_k|^4] + 
		\mathbb{E}[|s_k|^2] }{4\mathbb{E}[|s_k|^2] -2 } \text{Tr}(\mathbf{HH^H}) +\\& ||\mathbf{w'}^H_{zf,L} \mathbf{H'}||^2 +\sigma^2_v \mathbf{e}^H_D \mathbf{(HH^H)}^{-1} \mathbf{e}_D.
	\label{eq:error_var3}
	\end{align}
The probability  $\Pr(\mathcal{M}_{\hat k}=\mathcal{M}_l | \mathcal{M}_k, \mathbf{h})$ is a function of $f(s_k +\epsilon)$ and is computed by evaluating \cite[Eqn. 10]{urriza2011} for SNR $\gamma=1/\sigma^2_\epsilon$. The expression is not repeated here due to lack of space. Finally, the probability of correct classification is computed as $P_c(\mathbf{h}) = \sum_{k=1}^{K}\Pr(\mathcal{M}_{\hat k}=\mathcal{M}_k | \mathcal{M}_k,\mathbf{h}) \Pr(\mathcal{M}_k)$, where $\Pr(\mathcal{M}_k)$ is the probability that modulation level $\mathcal{M}_k$ is transmitted.

\begin{figure}
	\centering
	\begin{subfigure}[b]{0.22 \textwidth}
		\centering
		\includegraphics[width= \columnwidth]{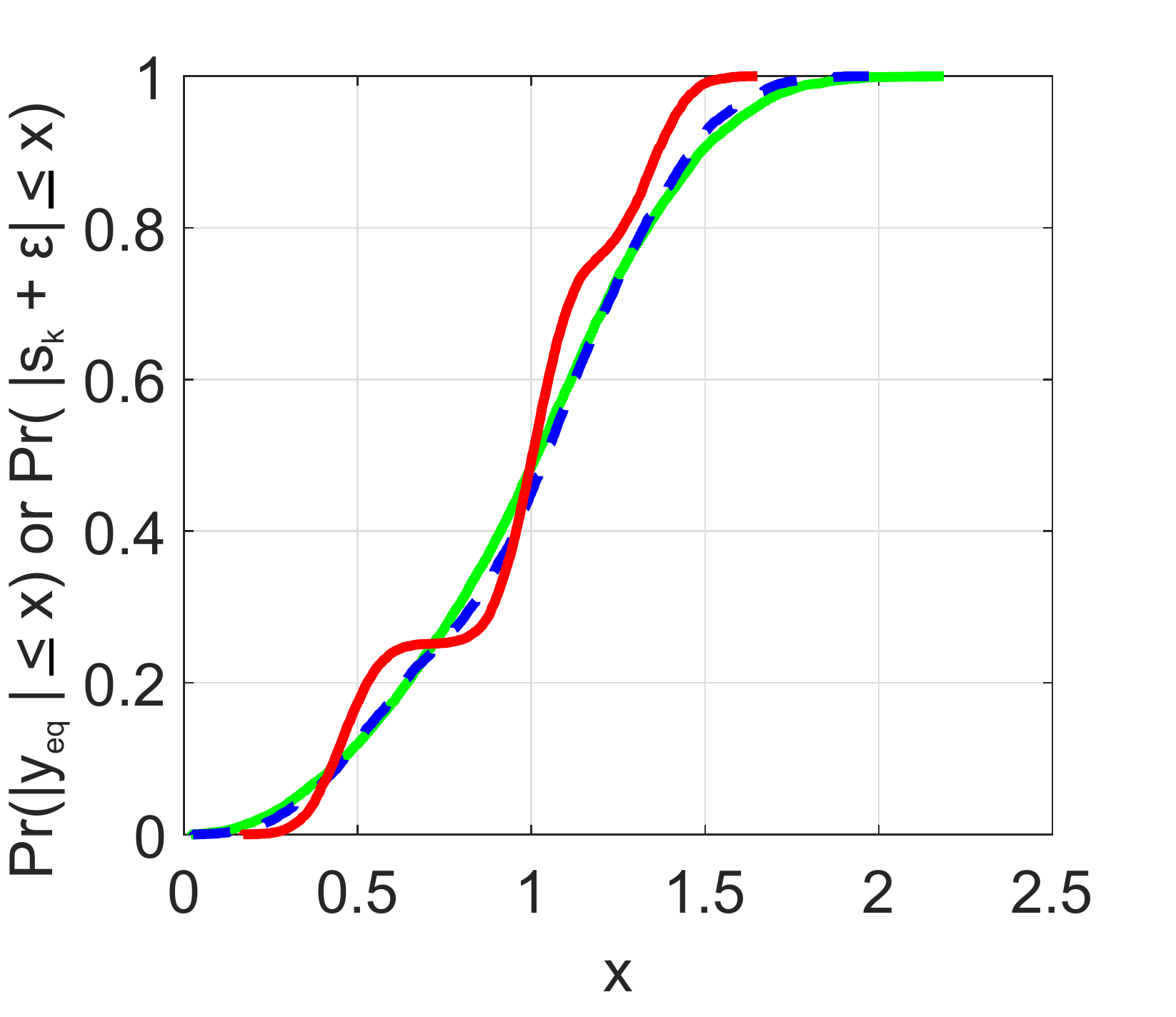}
		\caption{SNR: 20dB.}
	\end{subfigure}		
	~
	\begin{subfigure}[b]{0.25 \textwidth}
		\centering
		\includegraphics[width= \columnwidth]{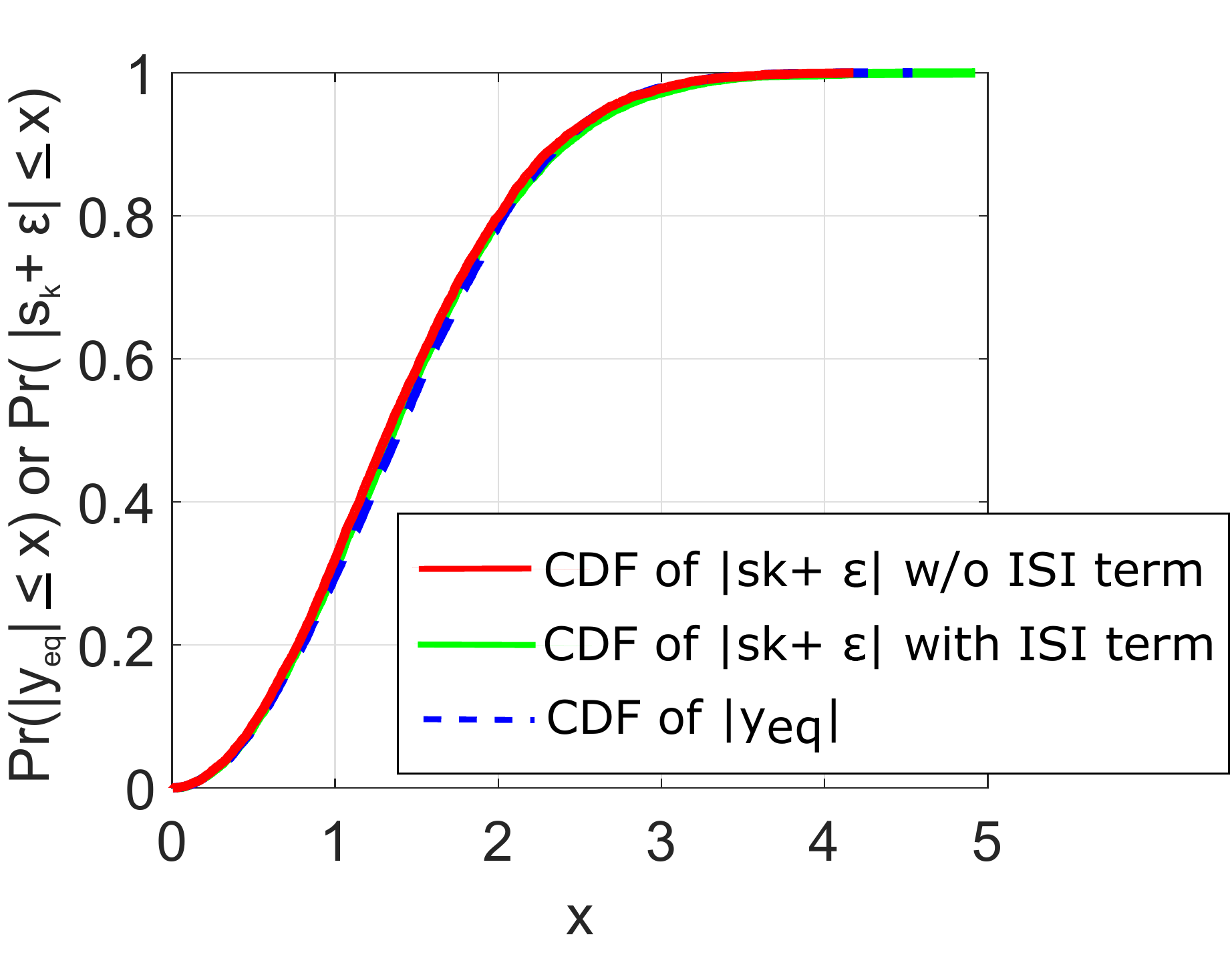}
		\caption{SNR: 0dB.}
	\end{subfigure}	
	\caption{\footnotesize CDFs of $F_{y_{eq}}=|y_{eq}|$ and $|s_k +\epsilon|$ with and without the residual ISI term $||\mathbf{w^{'H}_{zf,L} H'}||^2$. $\mathcal{M}_k$=16-QAM, Channel: $\mathbf{h}= [1,0,0.9]^T$. $L_{zf}=90$, $L=20$, and $\mu=10^{-4}$.} 
	\label{fig:cdfs_1009}
	\vspace{-4mm}
\end{figure}

\section{Simulation Results}
\label{sec:Results}
\begin{figure*}
	\centering
	\begin{subfigure}[b]{0.3 \textwidth}
		\centering
		\includegraphics[width= \columnwidth]{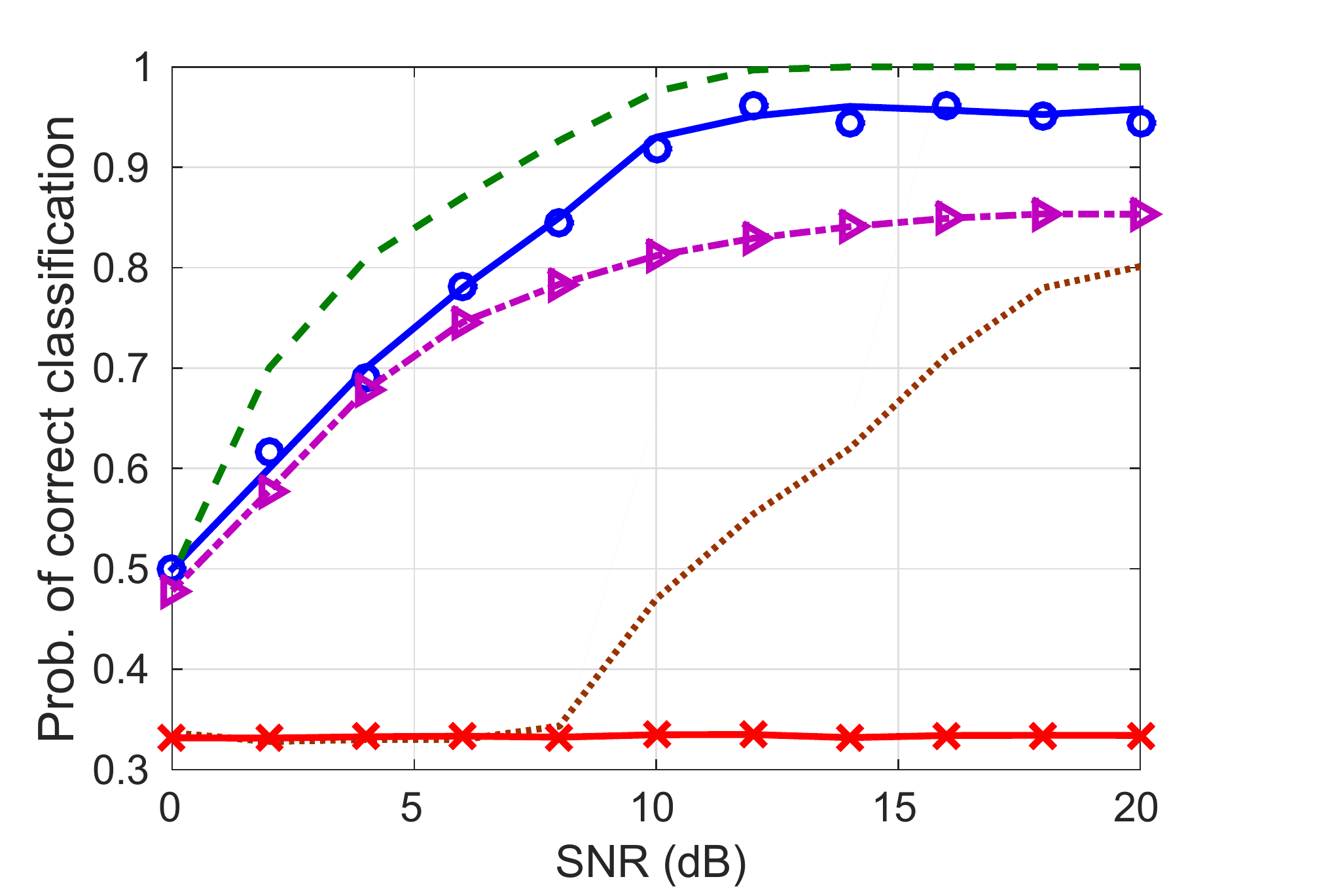}
		\caption{\footnotesize \{4, 16, 64\}-QAM classification under ch-1.}
		\label{fig:qam_ch1}
	\end{subfigure}		
	~
	\begin{subfigure}[b]{0.3 \textwidth}
		\centering
		\includegraphics[width= \columnwidth]{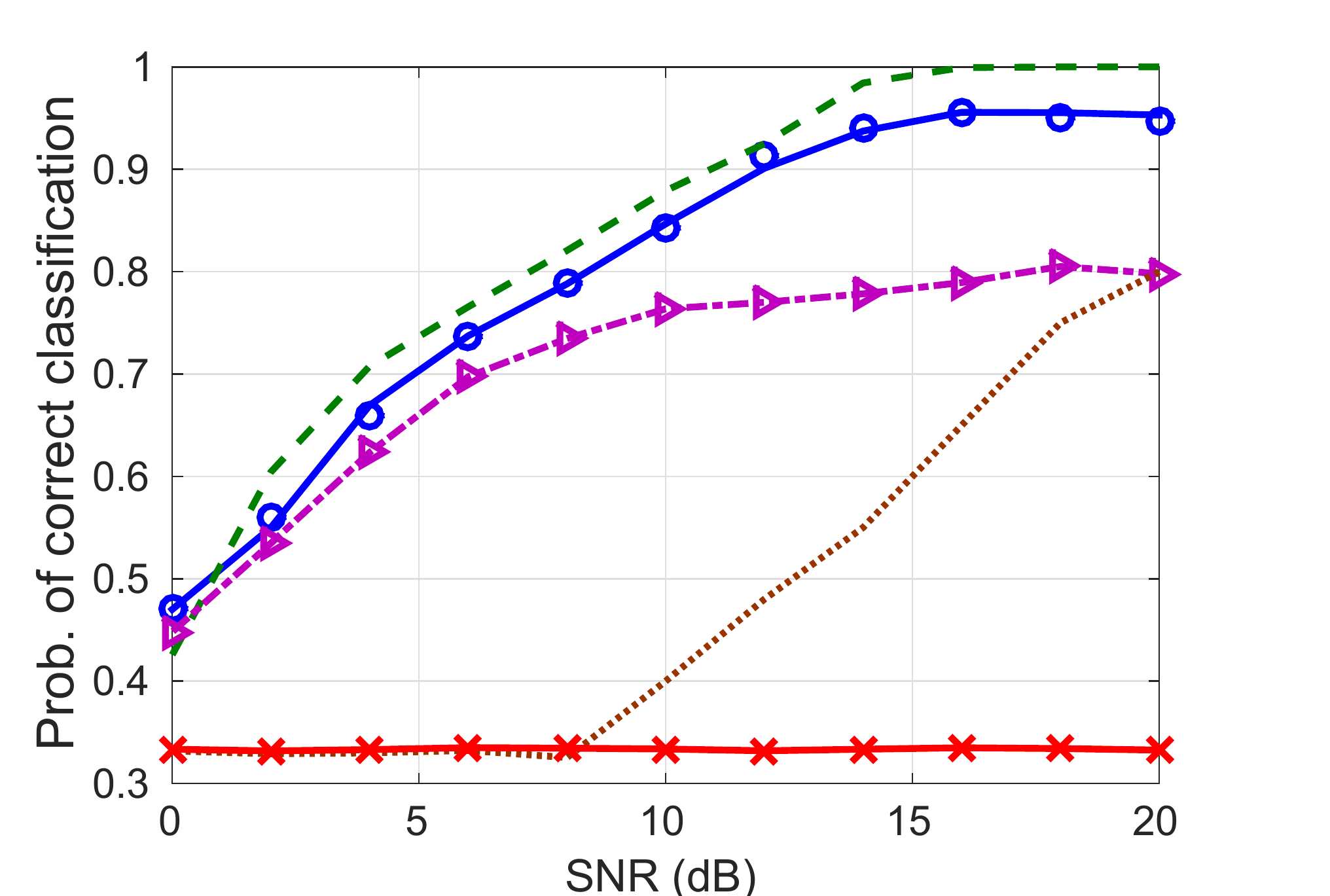}
		\caption{\footnotesize \{4, 16, 64\}-QAM classification under ch-2.}
		\label{fig:qam_ch2}
	\end{subfigure}			
	~
	\begin{subfigure}[b]{0.3 \textwidth}
		\centering
		\includegraphics[width= \columnwidth]{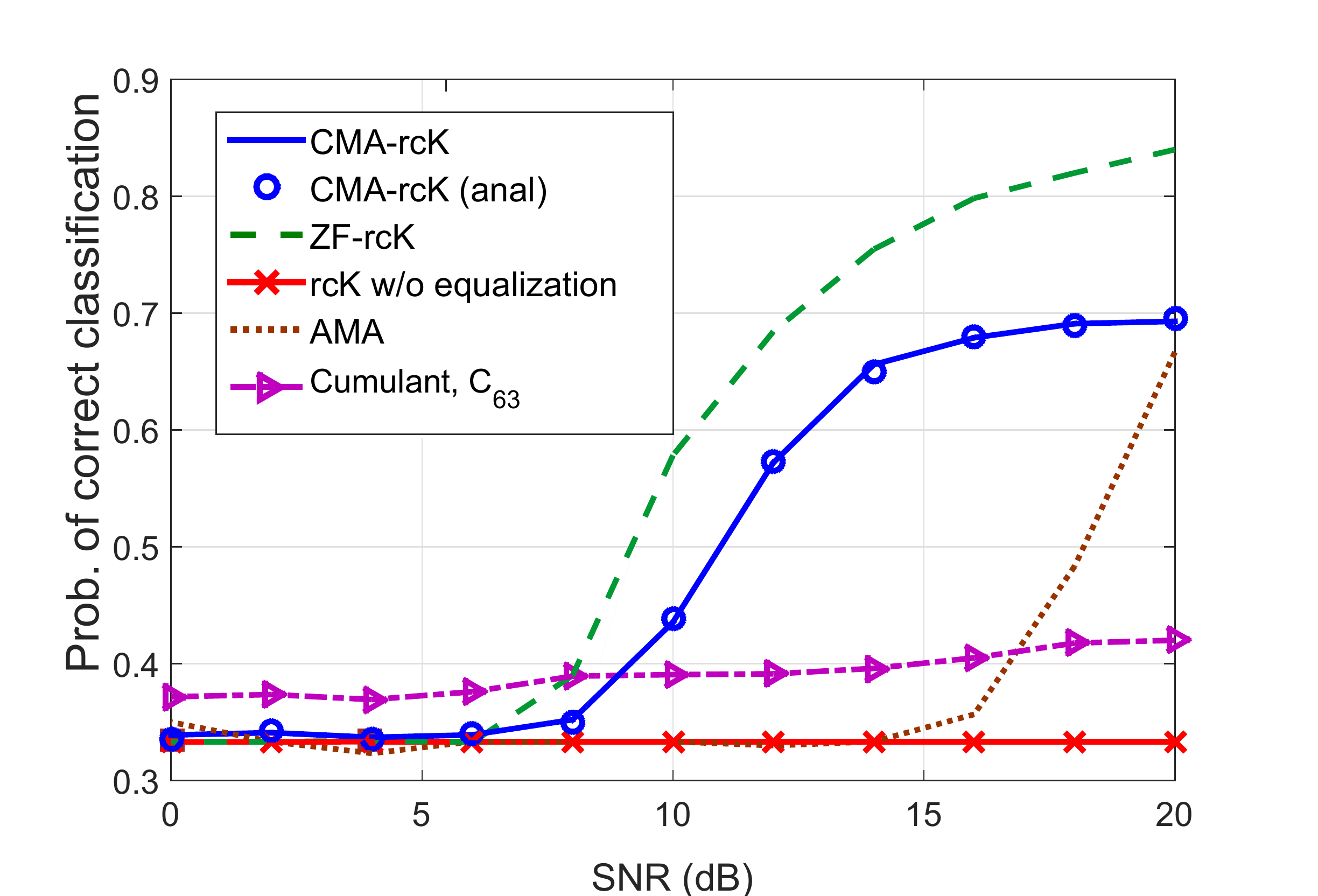}
		\caption{\footnotesize \{4, 16, 64\}-QAM classification under ch-3.}
		\label{fig:qam_ch3}
	\end{subfigure}	
	~
	\begin{subfigure}[b]{0.3 \textwidth}
		\centering
		\includegraphics[width= \columnwidth]{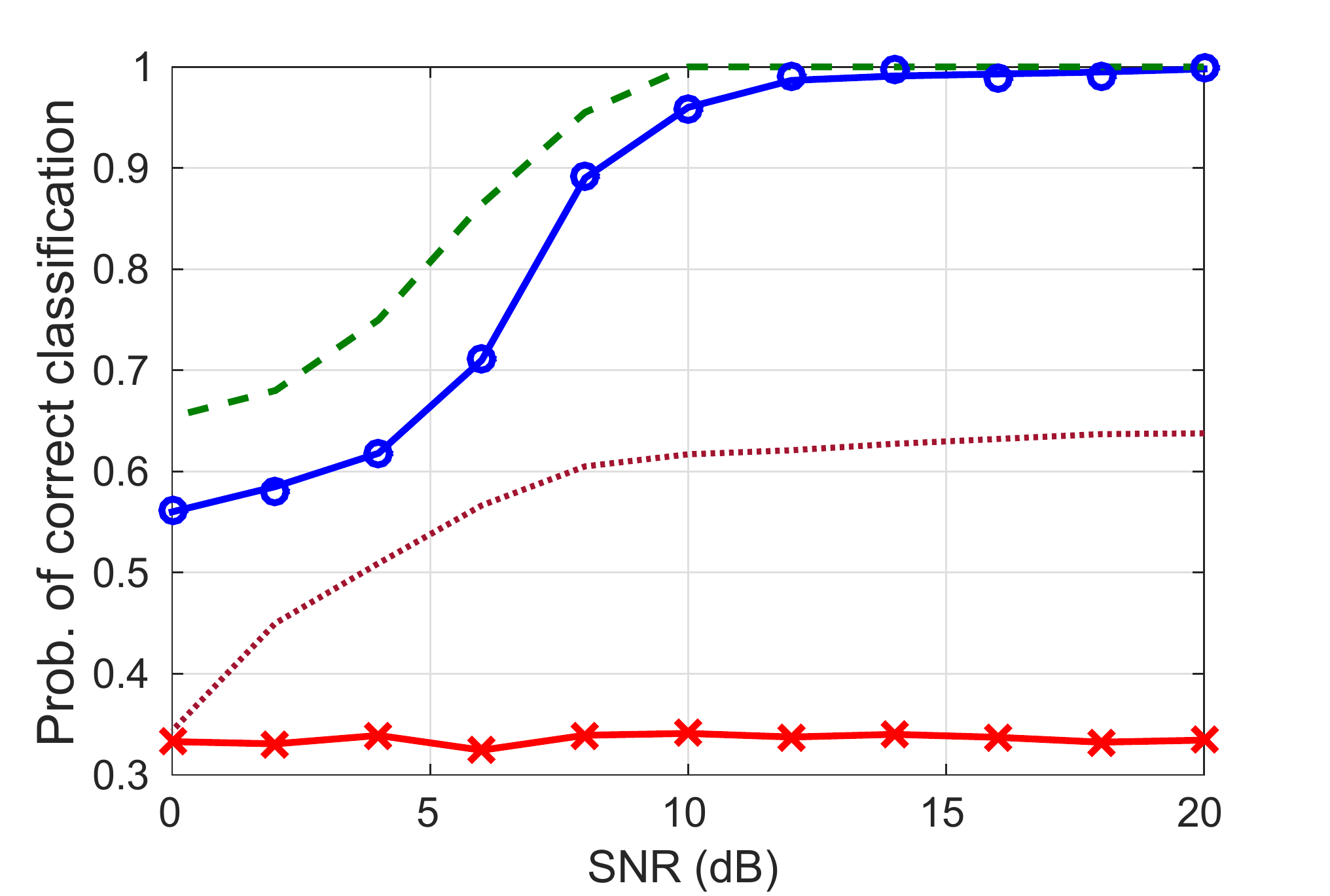}
		\caption{\footnotesize \{2, 4, 8\}-PSK classification under ch-1.}
		\label{fig:psk_ch1}
	\end{subfigure}				
	~
	\begin{subfigure}[b]{0.3 \textwidth}
		\centering
		\includegraphics[width= \columnwidth]{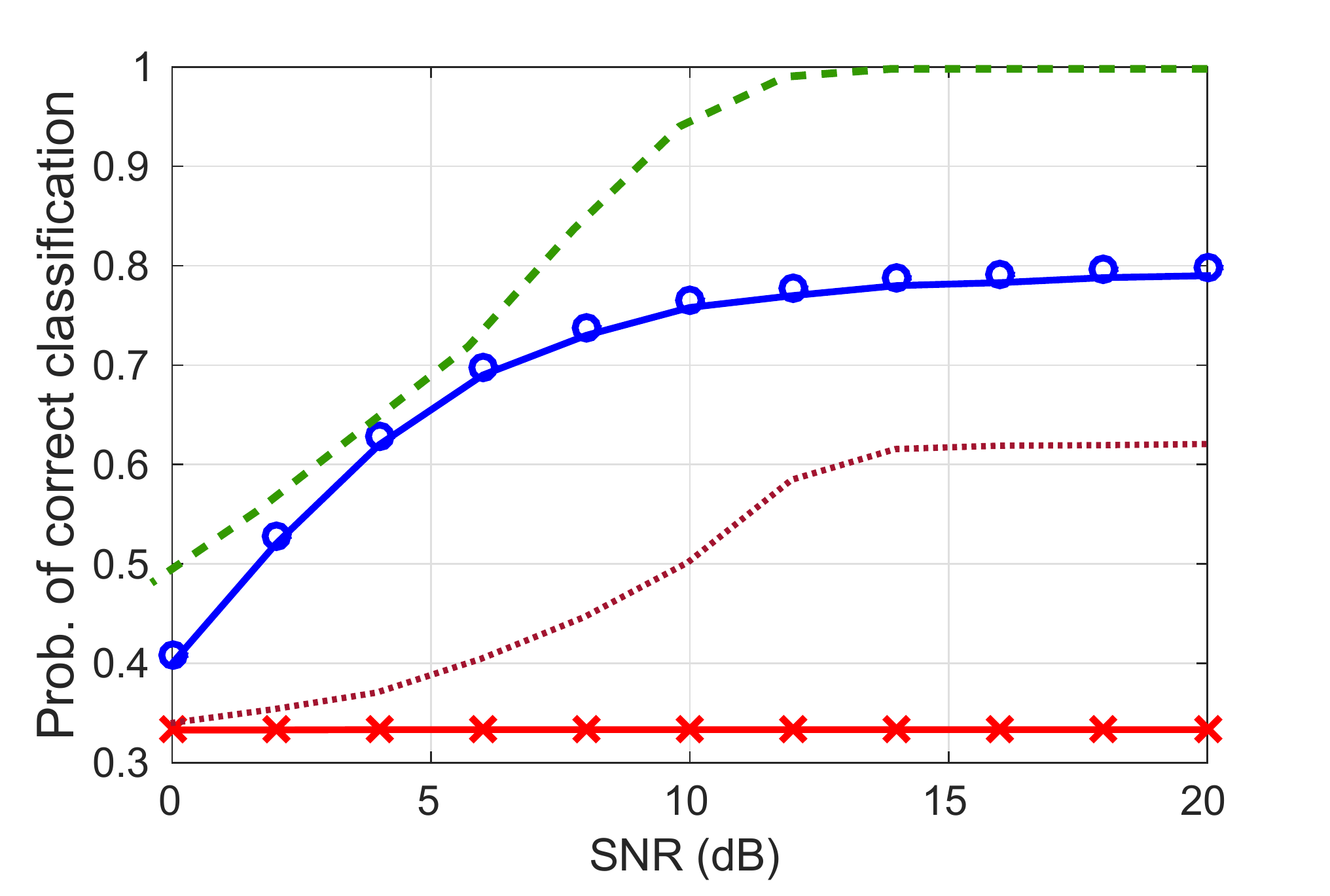}
		\caption{\footnotesize \{2, 4, 8\}-PSK classification under ch-2.}
		\label{fig:psk_ch2}
	\end{subfigure}				
	~
	\begin{subfigure}[b]{0.3 \textwidth}
		\centering
		\includegraphics[width= \columnwidth]{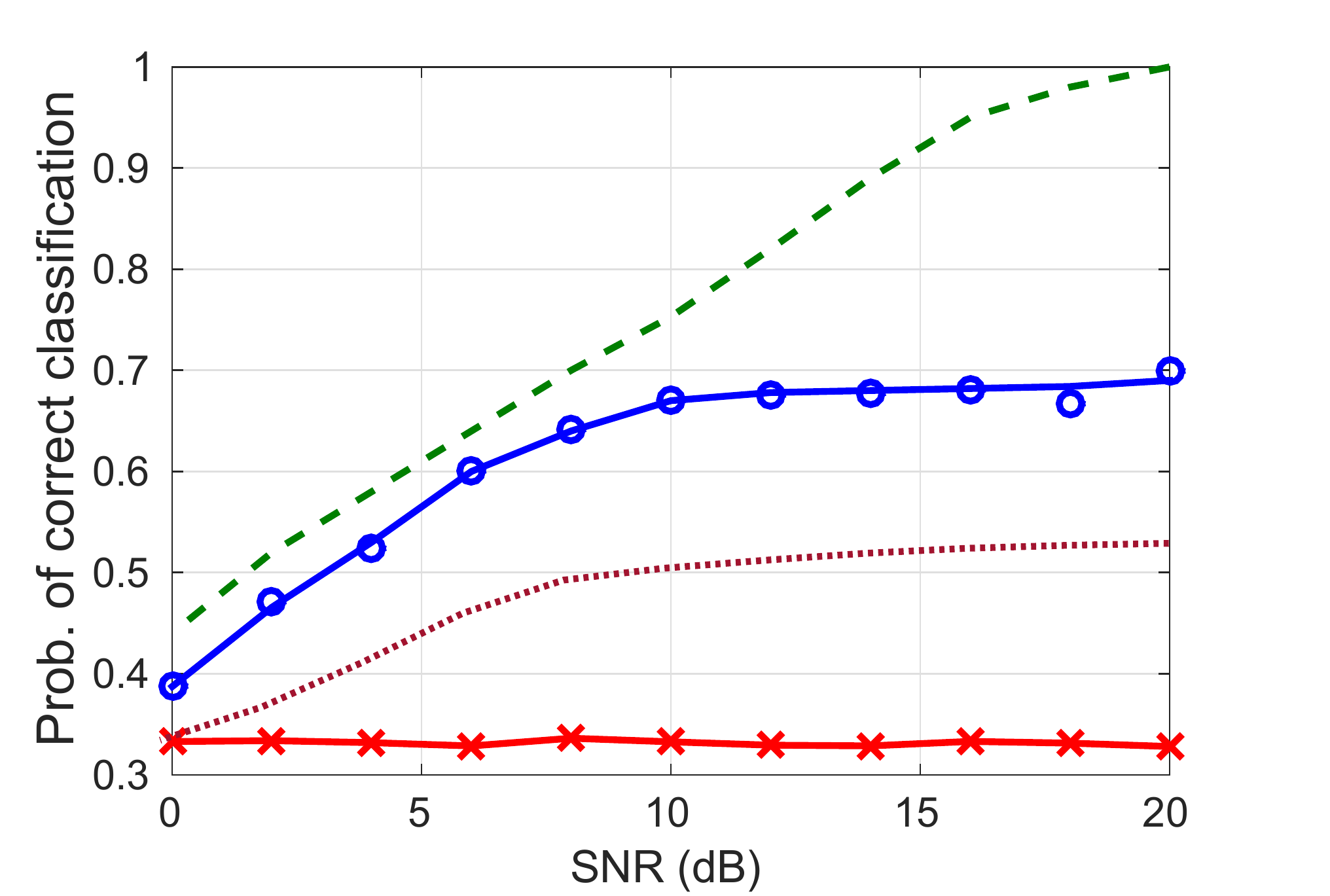}
		\caption{\footnotesize \{2, 4, 8\}-PSK classification under ch-3.}
		\label{fig:psk_ch3}
	\end{subfigure}				
	\caption{\footnotesize $P_c$ vs SNR for QAM and PSK signals. $L=20$, $M=200$. Legends in all figures are same as in Fig. \ref{fig:qam_ch3}.}
	\label{fig:all_results}
	\vspace{-6mm}
\end{figure*}

The distribution of $f(s_k+\epsilon)$ is a function of $\sigma^2_\epsilon$, which in turn depends on the channel $\mathbf{h}$. For a highly frequency selective channel, we have $L_{zf} > L$ and the term $||\mathbf{w^{'H}_{zf,L}}\mathbf{H'}||^2$ becomes prominent. In order to show the importance of including this term in the variance of $\epsilon$, we plot the CDFs of $f(s_k +\epsilon)=|s_k+\epsilon|$ and $f(y_{eq})=|y_{eq}|$ for a 16-QAM signal received under a highly frequency selective channel: $\mathbf{h}=[1,0,0.9]$ as shown in Fig. \ref{fig:cdfs_1009}. It can be observed that the distribution of $|s_k+\epsilon|$ matches that of $|y_{eq}|$ only when the residual ISI term is included in the computation of $\sigma^2_\epsilon$. The difference in the distributions with and without including the ISI term is more prominent at higher SNR because the ISI term dominates the noise enhancement at higher SNR. 

The performance of the classifier is evaluated under three different channel models.
 Channel model 1 (ch-1) is a 4-tap Rayleigh fading channel used in \cite{orlic2010, wu2008} with $h(0)=1$ and $h(1), h(2), h(3)\sim \mathcal{CN} (0,0.05)$. Channel model 2 (ch-2) is a LTE channel model with tap variances obtained by sampling the Extended Vehicular A (EVA) model at symbol rate $1$MHz \cite{LTE_36_101}. It is a three tap model where $h(0)\sim  \mathcal{CN} (0,0.95), h(1)\sim  \mathcal{CN} (0,0.28), h(2)\sim  \mathcal{CN} (0,0.11)$. Channel model 3 (ch-3) has constant taps: $\mathbf{h}=[1,0,0.9]$. Block fading is assumed in ch-1 and ch-2 where the taps remain constant for $M=200$ iterations required in CMA iterations. The length of equalizer is $L=20$ and the total number of samples used in all classifiers is 4000.  The transmitted symbols $s_k$ are selected uniformly from symbol set $\mathcal{M}_{k}\in \{4,16,64\}$ for QAM level classification and $\mathcal{M}_{k}\in \{2,4,8\}$ for PSK level classification. The average probability of correct classification $P_c$ is computed by running simulations for 10000 realizations for each value of $\mathbf{h}$ generated according the three channel models. 
 The classification performance of CMA-rcK is compared with cumulant-based classifier and AMA classifier \cite{barbarossa2000} for QAM level classification.  It has been observed in \cite{orlic2010} that the classifier based on the sixth-order cumulant outperforms the classifier based on the forth-order cumulant. Therefore, we use the sixth-order cumulant, $C_{63}$, for comparison. Additionally, we present results with ZF-rcK classifier where symbols are equalized with ZF equalizer, i.e., $\mathbf{w_{zf}}^H \mathbf{x}_i$ are used in rcK. ZF-rcK requires the knowledge of channel taps, while other techniques (CMA-rck, AMA, and cumulants) do not require any channel knowledge. For PSK level classification, the performance of CMA-rcK is compared with AMA and ZF-rcK classifiers only, since cumulant-based classifier cannot distinguish between 4- and 8-PSK.

As shown in Fig. \ref{fig:all_results}, the classification performance heavily depends on the channel model used. The accuracy of all the classifiers under ch-1 is higher as compared to ch-2 and ch-3. This is because ch-1 generates taps with low frequency selectivity due to lower tap variances as compared to ch-2. Performance of the all the classifiers remains below 70\% under ch-3 due to its high frequency selectivity. This is because z-transform of the channel $\mathbf{h}=[1,0,0.9]$ has zeros close to the unit circle causing large ISI. The CMA equalizer is not able to completely remove the impact of ISI, thereby resulting in a low classification accuracy for both QAM and PSK level classification as seen in Fig. \ref{fig:qam_ch3} and \ref{fig:psk_ch3}, respectively. Howver, the proposed classifier achieves $>90$\% accuracy at 20dB SNR for QAM level classification under ch-1 and ch-2, and for PSK classification under ch-1 and it outperforms existing classifiers based on cumulants and AMA equalizer.

\vspace{-2mm}
\section{Conclusion}
\label{sec:Conclusion}
In this paper, we proposed a modulation level classification technique for the signal received under multipath channel without any prior knowledge of the channel. The proposed technique includes a blind equalizer using constant modulus algorithm (CMA), followed by level classification using reduced complexity Kuiper (rcK) classifier. The expression for the variance of residual error at the output of the equalizer is derived for a given multipath channel. The variance is then used to compute the probability of correct classification for the proposed classifier. It has been observed that the classification accuracy of the classifiers heavily depends on the frequency selectivity of the multipath channel. The numerical results show that the proposed CMA-rcK classifier outperforms existing methods based for QAM and PSK level classification under different channel models for SNRs between 0 and 20dB.

\bibliographystyle{IEEEtran}
\bibliography{IEEEabrv,references_v2, LTE_standard_citation}

\end{document}